\documentclass[fleqn,twoside]{article}

\usepackage{espcrc2}

\usepackage{graphicx}
\newcommand{\AmS}{{\protect\the\textfont2
  A\kern-.1667em\lower.5ex\hbox{M}\kern-.125emS}}
\vspace{-2pc}
\title{\center{ACORDE a Cosmic Ray Detector for ALICE}}
\author{\center{
\vspace{1pc}
}
A.~Fern\'andez\address[PUE]{Benem\'erita Universidad Autonoma de
Puebla, Puebla, Mexico},
E.~G\'amez\addressmark[PUE],
G.~Herrera\address[A]{Depto. Fisica, Centro de Investigacion y de
Estudios Avanzados, Mexico, Mexico},
R.~L\'opez\addressmark[PUE], I.~Le\'on-Monz\'on\addressmark[PUE]
M.~I.~Mart\'{\i}nez\address[C]{Inst. Ciencias Nucleares, Universidad
Nacional Autonoma de Mexico, Mexico, Mexico},
C.~Pagliarone\address[PIS]{Universit\'a di Cassino and Istituto Nazionale di Fisica Nucleare sez. Pisa, Italy}\thanks{Corresponding author. E-mail: pagliarone@fnal.gov},\\
G.~Paic\addressmark[C],
S.~Rom\'an\addressmark[PUE],
G.~Tejeda\addressmark[PUE],
M.~A.~Vargas\addressmark[PUE],
S.~Vergara\addressmark[PUE],
L.~Villase\~nor\address{Inst. Fisica y Matematicas, Universidad
Michoacana de San Nicolas de Hidalgo, Morelia, Mexico},
A.~Zepeda\addressmark[A]\\
(The ACORDE Collaboration)
}
\begin{document}
%%%%%%%%%%%%%%%%%%%%%%%%%%%%%%%%%%%%%%%%%%%%%%%%%%%%%%%%%%%%%%%%%%%%%%%%%%%%%%%%%%%%%%%%%%%%%%%%%
\vspace{2pc}
\begin{abstract}
ACORDE is one of the ALICE detectors, presently under construction
at CERN. It consists of an array of plastic scintillator counters
placed on the three upper faces of the ALICE magnet. It will act as
a cosmic ray trigger, and, together with other ALICE sub-detectors,
will  provide precise information on cosmic rays with primary
energies around $10^{15} \div 10^{17}$ eV. Here we describe the
design  of ACORDE along with the present status and integration into
ALICE.
\end{abstract}
%%%%%%%%%%%%%%%%%%%%%%%%%%%%%%%%%%%%%%%%%%%%%%%%%%%%%%%%%%%%%%%%%%%%%%%%%%%%%%%%%%%%%%%%%%%%%%%%%

%%%%%%%%%%%%%%%%%%%%%%%%%%%%%%%%%%%%%%%%%%%%%%%%%%%%%%%%%%%%%%%%%%%%%%%%%%%%%%%%%%%%%%%%%%%%%%%%%
\maketitle
\section{Introduction}
%%%
%The question whether the knee of the cosmic-ray energy spectrum, around $3 \times 10^{15} eV$,
%has an astrophysical origin or whether it is due to a change in the properties of the hadronic
%interactions, is an important open question.
%%%
%Because of their extremely low flux, the study of cosmic ray particles with energies around and
%beyond this knee region ($10^{15} \div 10^{-17}$ eV), is possible only using indirect methods.
%%%
%%%%%%%%%%%%%%%%%%%%%%%%%%%%%%%%%%%%%%%%%%%%%%%%%%%%%%%%%%%%%%%%%%
%%%

At CERN, the use of large underground high-energy physics experiments, for comic ray studies,
has an important tradition~\cite{add2}.
%%%
%%%%%%%%%%%%%%%%%%%%%%%%%%%%%%%%%%%%%%%%%%%%%%%%%%%%%%%%%%%%%%%%%%
ALICE is an experiment mainly designed to study the products of
nucleus-nucleus collisions at CERN Large Hadron Collider
(LHC)~\cite{ppr}.
%%%
%%%%%%%%%%%%%%%%%%%%%%%%%%%%%%%%%%%%%%%%%%%%%%%%%%%%%%%%%%%%%%%%%%
The underground location of the ALICE detector, with 30\,m of
overburden composed of subalpine molasse, is an ideal place for muon
based underground experiments. Using this facilities, we plan to
observe muon bundles generated by cosmic ray primary particles with
energies around the knee region $10^{15} \div
10^{17}$eV~\cite{vienna}.
%%%
%%%%%%%%%%%%%%%%%%%%%%%%%%%%%%%%%%%%%%%%%%%%%%%%%%%%%%%%%%%%%%%%%%
%%%
ACORDE (the ALICE COsmic Ray DEtector) is an array of scintillator
modules that will act as  a cosmic ray trigger for ALICE
calibration, as well as, multiple muon trigger to study high energy
cosmic rays.

%detect atmospheric muons and multi-muons events (muon bundles). %%
%%
%% \vspace{1pc}
%% \par\noindent{\bf\underline{The ACORDE Detector}}\\
%%
%%
\vspace{-0.5pc}
\section{The ACORDE Detector}
ACORDE is an array of plastic scintillator modules (60 at the present)
placed on the top sides of the central ALICE magnet, as shown in
Figure~1. More modules, to achieve a better angular coverage and
acceptance, will be added later.
Each module, see Figure~2, consists of two plastic scintillator paddles  with
$188\times 20$~cm$^2$ effective area, arranged in a doublet configuration.
Each doublet consists of two superimposed scintillator counters, with their
corresponding photomultipliers active faces, looking back to back.
A coincidence signal, in a time window of 40 ns, from the two scintillator
paddles gives, for each module, the trigger hit.
A PCI BUS electronics card have been developed in order to measure plateau
and efficiency of the module counters~\cite{scint}.
The signal of each ACORDE scintillator channel is applied to a
leading edge discriminator. After the conversion of each PMT negative signal in a digital hit,
a single or a multi-coincidence trigger signal is generated.
In both cases we will use a tracking system to identify which ACORDE modules were hit.
Afterwards, these module hit information will be stored in a FIFO memory before being
sent to the DAQ system through the Detectors Data Link (DDL).
%%%%%%%%%%%%%%%%%%%%%%%%%%%%%%%%%%%%%%%%%%%%%%%%%%%%%%%%%%%%%%%%%%%%%%%%%%%%%%%%%%%%%%%%%%

%%%%%%%%%%%%%%%%%%%%%%%%%%%%%%%%%%%%%%%%%%%%%%%%%%%%%%%%%%%%%%%%%%%%%%%%%%%%%%%%%%%%%%%%%%%%%%%%%%%%%%
\begin{figure}[t!]
\begin{center}
\includegraphics[width=4.8cm,height=4.8cm]{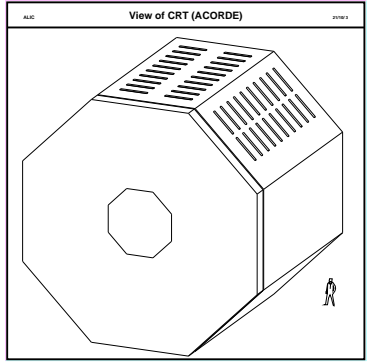}
%\end{center}
\vspace{-1pc}
\caption{\footnotesize The ALICE Cosmic Ray Detector. The scintillator array is on top of the ALICE magnet.}
\label{DETcrt_new-array}
\end{center}
\vspace{-2pc}
\end{figure}

%%%%%%%%%%%%%%%%%%%%%%%%%%%%%%%%%%%%%%%%%%%%%%%%%%%%%%%%%%%%%%%%%%%%%%%%%%%%%%%%%%%%%%%%%%%%%%%%%%%%%%

%%%%%%%%%%%%%%%%%%%%%%%%%%%%%%%%%%%%%%%%%%%%%%%%%%%%%%%%%%%%%%%%%%%%%%%%%%%%%%%%%%%%%%%%%%%%%%%%%
\section{The ALICE Cosmic Ray Trigger}

The cosmic ray trigger system (CRT) will provide a fast level-zero
trigger signal to the central trigger processor, when atmospheric
muons impinge upon the ALICE detector. The signal will be useful for
calibration, alignment and performance of several ALICE tracking
detectors, mainly the ALICE Time Projection Chamber (TPC), the ALICE
Transition Radiation Detector (TRD)  and the ALICE Inner Tracking
System (ITS). The typical rate for single atmospheric muons crossing
the ALICE cavern will be less than  Hz/m$^2$. The expected rate for
multi-muon events will be around or less than $0.04$ Hz/m$^2$.
Atmospheric muons will need an energy of at least $17$ GeV to reach
the ALICE hall, while the upper energy limit for reconstructed muons
in the TPC will be less than 2 TeV, depending on the ALICE magnetic
field intensity (up to $0.5$ T). We have designed and implemented
the necessary electronics to do the following tasks: LHC clock
synchronization, send single and multi-muon trigger signal to the
Central Trigger processor, send a wake-up signal to ALICE-TRD and
communicate with the DAQ through a DAQ Source Interface card(SIU).
Also, we complete the ACORDE Detector Control System, to monitor the
performance of the scintillator counter array.

%%%%%%%%%%%%%%%%%%%%%%%%%%%%%%%%%%%%%%%%%%%%%%%%%%%%%%%%%%%%%%%%%%%%%%%%%%%%%%%%%%%%%%%%%%%%%%%%%
\begin{figure}[t!]
\begin{center}
\includegraphics[width=7cm,height=4.8cm]{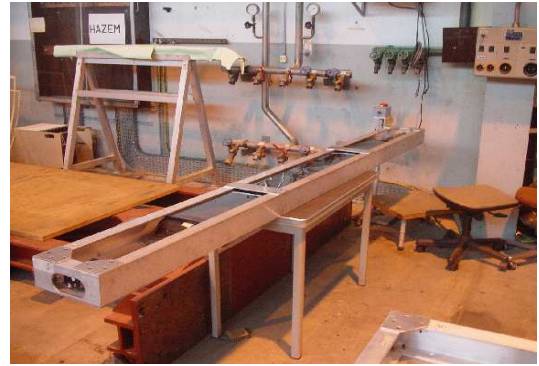}
\vspace{-1pc} \caption{\footnotesize An ACORDE module; the
scintillator modules consist of two superimposed scintillator
counters.} \label{ang-dist}
\end{center}
%\vspace{-2.86pc}
\vspace{-1.1pc}
\end{figure}
%%%%%%%%%%%%%%%%%%%%%%%%%%%%%%%%%%%%%%%%%%%%%%%%%%%%%%%%%%%%%%%%%%%%%%%%%%%%%%%%%%%%%%%%%%%%%%%%%

At the present, we have 20 ACORDE modules already installed and the
related electronics working. The ALICE-TPC above ground
commissioning is proceeding  based on  ACORDE trigger using 10
modules placed on the top and 10 underneath the ALICE TPC
(see~\cite{glassel}).
%%%%%%%%%%%%%%%%%%%%%%%%%%%%%%%%%%%%%%%%%%%%%%%%%%%%%%%%%%%%%%%%%%%%%%%%%%%%%%%%%%%%%%%%%%%%%%%%%

\section{Summary}

We have implemented a dedicated cosmic ray trigger for ALICE  which
in conjunction with other ALICE detectors provides a powerful tool
for the study of muon bundles properties.

%%  \section*{Acknowledgments}
%%
%%  The author thanks the organizing committee of the conference for their kindness.
%%
%%%%%%%%%%%%%%%%%%%%%%%%%%%%%%%%%%%%%%%%%%%%%%%%%%%%%%%%%%%%%%%%%%%%%%%%%%%%%%%%%%%%%%%%%%%%%%%%%

%%%%%%%%%%%%%%%%%%%%%%%%%%%%%%%%%%%%%%%%%%%%%%%%%%%%%%%%%%%%%%%%%%%%%%%%%%%%%%%%%%%%%%%%%%%%%%%%%

\end{document}